# NR Conformance Testing of Analog Radio-over-LWIR FSO Fronthaul link for 6G Distributed MIMO Networks

**Rafael Puerta[1,2], Mengyao Han[2,4], Mahdieh Joharifar[2], Richard Schatz[2], Yan-Ting Sun[2], Yuchuan Fan[2,3], Anders Djupsjöbacka[3], Grégory Maisons[5], Johan Abautret[5], Roland Teissier[5], Lu Zhang[6], Sandis Spolitis[7], Muguang Wang[4], Vjaceslavs Bobrovs[7], Sebastian Lourdudoss[2], Xianbin Yu[6], Sergei Popov[2], Oskars Ozolins[3,2,7], and Xiaodan Pang[2,3,7]**

*[1]Ericsson Research, Ericsson AB, 164 83 Stockholm, Sweden, rafael.puerta@ericsson.com*
*[2]Department of Applied Physics, KTH Royal Institute of Technology, 106 91 Stockholm, Sweden, xiaodan@kth.se*
*[3]RISE Research Institutes of Sweden, 16440 Kista, Sweden, oskars.ozolins@ri.se*
*[4]Institute of Lightwave Technology, Key Lab of All Optical Netw. & Advanced Telecom Netw., EMC, Beijing Jiaotong University, Beijing, China*
*[5]mirSense, 2 Bd Thomas Gobert 91120 Palaiseau, France*
*[6]College of Information Science and Electronic Engineering, Zhejiang University, and Zhejiang Lab, Hangzhou, China*
*[7]Institute of Telecommunications, Riga Technical University, 1048 Riga, Latvia*

**Abstract:** We experimentally test the compliance with 5G/NR 3GPP technical specifications of an analog radio-over-FSO link at 9 µm. The ACLR and EVM transmitter requirements are fulfilled validating the suitability of LWIR FSO for 6G fronthaul. 

## 1. Introduction

5G is paving the way for the transformation and digitalization of key industry sectors like healthcare, smart cities, transportation, entertainment, agriculture, manufacturing, and others. However, it is challenging to simultaneously meet all the networking requirements since typical applications cover real-time high-resolution videos and holographic experiences demanding massive amounts of data, with uninterrupted service everywhere even at high mobility [1]. The key performance indicators (KPIs) of 5G for download and upload speeds are 100 Mbps and 50 Mbps, respectively [2]. However, the long-term requirements are substantially higher than that to fully support applications like the eXtended Reality (XR) and digital twins. Thus, new solutions that can provide higher data rates and a true ubiquitous service are needed for the next generation mobile networks, i.e., the 6G. In contrast to traditional deployments where the antennas are co-located in a single array in the center of the cell, the lately introduced distributed multiple-input multiple-output (D-MIMO) networks are equipped with geographically distributed antennas that can operate jointly, synchronously, and coherently to serve users. In such a way, D-MIMO can increase considerably the data rates and coverage, however, it is more complex and costly to deploy [3]. As D-MIMO requires synchronization of the fronthaul links of the spatially distributed antennas to realize coherent joint transmissions (CJTs), a cost-efficient solution to realize CJT is to use analog fronthaul links [4] in combination with centralized processing [5]. Free-space optical (FSO) links are a prospective alternative to fiber optics that fulfil the preceding requirements facilitating the deployment of D-MIMO networks. Recently, it has been shown that mid-infrared (IR) free-space optics (FSO) [6], specifically in the Long-wave IR (LWIR) band (8-12 µm), contain advantageous characteristics for terrestrial applications. It offers broader spectrum and lower atmospheric attenuation than the sub-THz/THz bands, whereas it is less sensitive to atmospheric turbulence and particle scattering than the traditional near-IR wavelengths used for fiber-optic telecom systems [7].

The 3rd Generation Partnership Project (3GPP) defines in technical specifications the radio frequency (RF) requirements and conformance test methods for New Radio (NR), i.e., 5G, comprising the transmitter and receiver characteristics [8]. A technology must satisfy all mandatory RF requirements to be considered NR compliant. In this paper, we experimentally validate the compliance with 5G/NR 3GPP technical specifications of analog radio-over-LWIR FSO fronthaul links. More specifically, we generate an NR signal with a bandwidth of 20 MHz and a carrier frequency of 627 MHz (NR band n71) and evaluate the 3GPP adjacent channel leakage power ratio (ACLR) and the error vector magnitude (EVM) transmitter requirements. To perform our tests, we used a 9.15-µm directly-modulated quantum cascade laser (DM-QCL) and a commercial mercury cadmium telluride (MCT) detector. In addition, we used 3GPP compliant RF test signals and procedures.

## 2. Experimental Setup

Figure 1(a) shows the block diagram of the experimental setup. An arbitrary waveform generator (AWG) with a resolution of 10-bits and a sampling frequency of 50 GSa/s generates the different 3GPP compliant NR test signals, i.e., test models (TMs). As baseline to all TMs, a bandwidth of 20 MHz with a subcarrier spacing (SCS) of 30 kHz, and carrier frequency of 627 MHz are used. This NR band is selected due to the limited bandwidth of the MCT detector. The DM-QCL is a 4 mm single mode distributed-feedback laser mounted epi-up on an Aluminum-Nitride submount, so that it can be operated at room temperature using a Peltier cooling module with a few tens of milliwatts of continuous wave infrared light at a wavelength of 9.15 µm. The DM-QCL chip is mounted on a commercial QCL mount, as shown in Fig. 1(b). The laser is biased and modulated through an external bias-tee. In this case, we operate the DM-QCL at 15°C to ensure sufficient incident power for the MCT detector. Fig. 1(c) shows the measured spectrum of the DM-QCL at 15°C. The P-I-V curve of the DM-QCL measured at 15°C is shown in Fig. 1(e). At this temperature, the lasing threshold is around 420 mA. When the laser bias current is up to 670 mA, the DM-QCL chip has an output power of 21 mW (well within the linear region). Fig. 1(b) and 1(d) show photos of the LWIR transmitter and receiver.

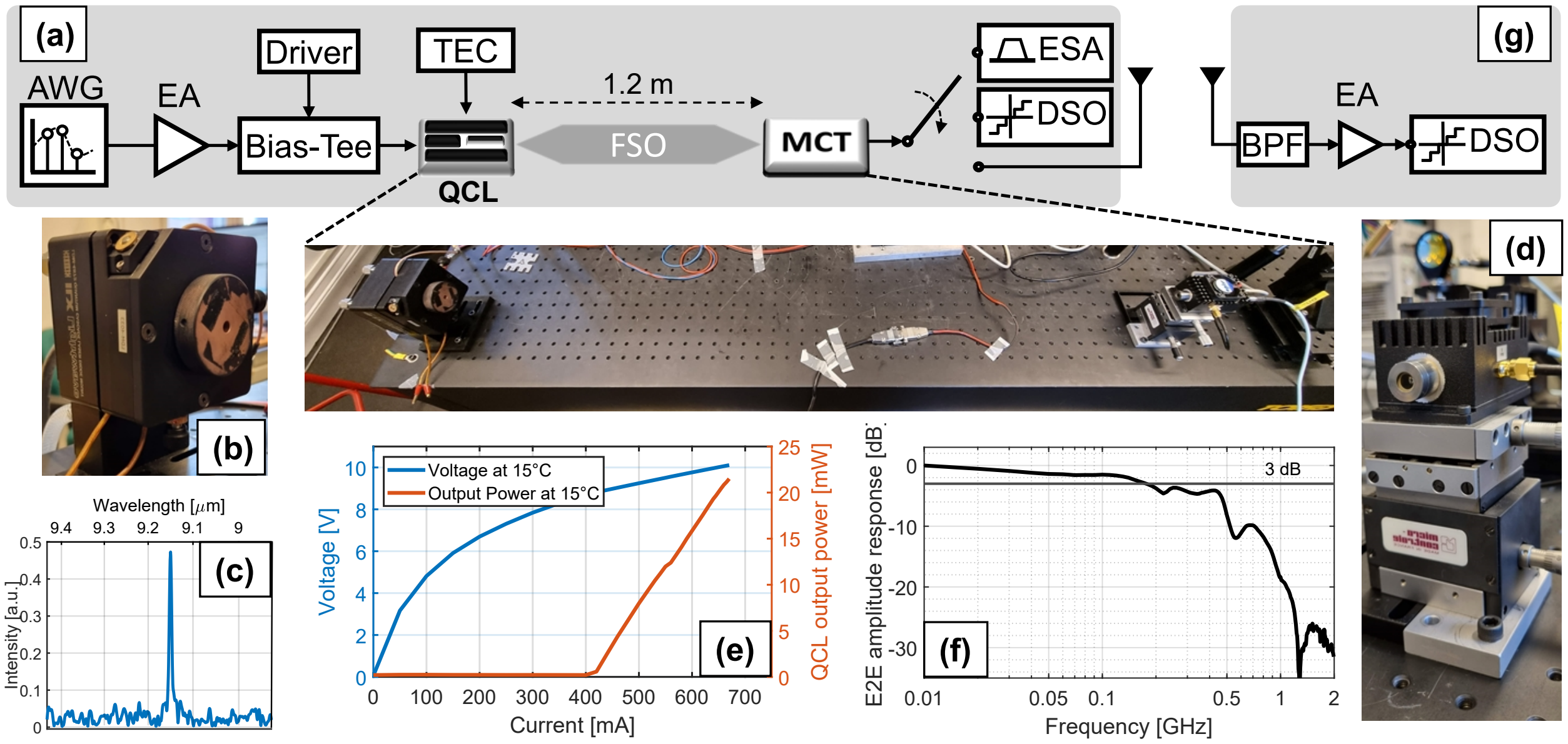

Fig. 1. (a) Experimental setup. (b) DM-QCL transmitter. (c) QCL output spectrum at room temperature. (d) MCT detector. (e) P-I-V curve of the 9.15-µm DM-QCL. (f) Characterized end-to-end amplitude response including the QCL, MTC detector, all the electrical and RF components. (g) Wireless receiver setup.

For the FSO transmission, we used a beam-collimating lens installed at the QCL mount to collimate the beam. The detector is a commercial MCT IR Photovoltaic (PV) detector. The wireless distance between the DM-QCL and the MCT detector is 1.2 m. The 3-dB bandwidth of the MCT detector is around 720 MHz. Fig. 1(f) shows the calibrated frequency response of the FSO channel, including the AWG, DM-QCL, MCT detector, and all the electrical components in between. We first calibrate the received signal power by using an IR power meter at the receiver side. Then, after the optimization of the received signal power, we replace the IR power meter with the MCT detector to perform the optical-to-electrical conversion (O/E). After O/E conversion, an electrical spectrum analyzer (ESA) is used to measure the ACLR of the RF signals using the parameters and procedure specified in [8]. Additionally, the RF signals are captured by a digital storage oscilloscope (DSO) with a resolution of 8-bits and a sampling frequency of 10 GSa/s for further offline processing. To calculate the EVM, first RF carrier recovery and signal downconversion is done by a digital Costas loop. Then, after cyclic prefix (CP) removal and time domain to frequency domain conversion, a zero-forcing (ZF) equalizer is applied as specified in [8]. Furthermore, for end-to-end (E2E) performance tests, Fig. 1(g) shows a wireless receiver setup comprising a multiband antenna, a narrow bandpass filter (BPF), and an EA.

## 3. Experimental Results

The following tests are done only for a single antenna connector and for a single NR carrier. To test the RF signals unwanted emissions requirement, i.e., ACLR, we used the 3GPP frequency range 1 (FR1, 410-7125 MHz) test signals NR-FR1-TM1.1 and NR-FR1-TM1.2. All subcarriers of these test signals carry QPSK symbols. The difference between these is that NR-FR1-TM1.2 assigns different power levels to different subcarriers. Fig. 2(a) and Fig. 2(b) show the electrical spectrum and the ACLR values of the RF signals after O/E and amplification, respectively. For both tests, the ACLR values of the adjacent channels are higher than 44.2 dB satisfying the 3GPP ACLR requirement [8]. It is noted that the bandwidth to calculate the ACLR is 18.36 MHz which corresponds to $N_{RB}$×SCS×12 with a SCS = 30 kHz and 51 resource blocks (RBs) [8].

To test the RF signals quality, i.e., EVM, we used the test signals NR-FR1-TM3.1 and NR-FR1-TM3.1a. All subcarriers of these signals carry 64-QAM and 256-QAM symbols respectively. Fig. 2(c) and Fig. 2(d) show the EVM values and received symbols constellations after O/E and amplification respectively. The physical downlink shared channel (PDSCH) root-mean square (RMS) EVM values are below 9% and 4.5% for 64-QAM and 256-QAM schemes respectively [8]. Moreover, the minimum requirements in [9] are also satisfied, i.e., 3.5% and 8% respectively. Also, it can be seen that some symbols (corresponding to the direct current subcarrier) have higher EVM due to the baseband downconversion process additional distortions.

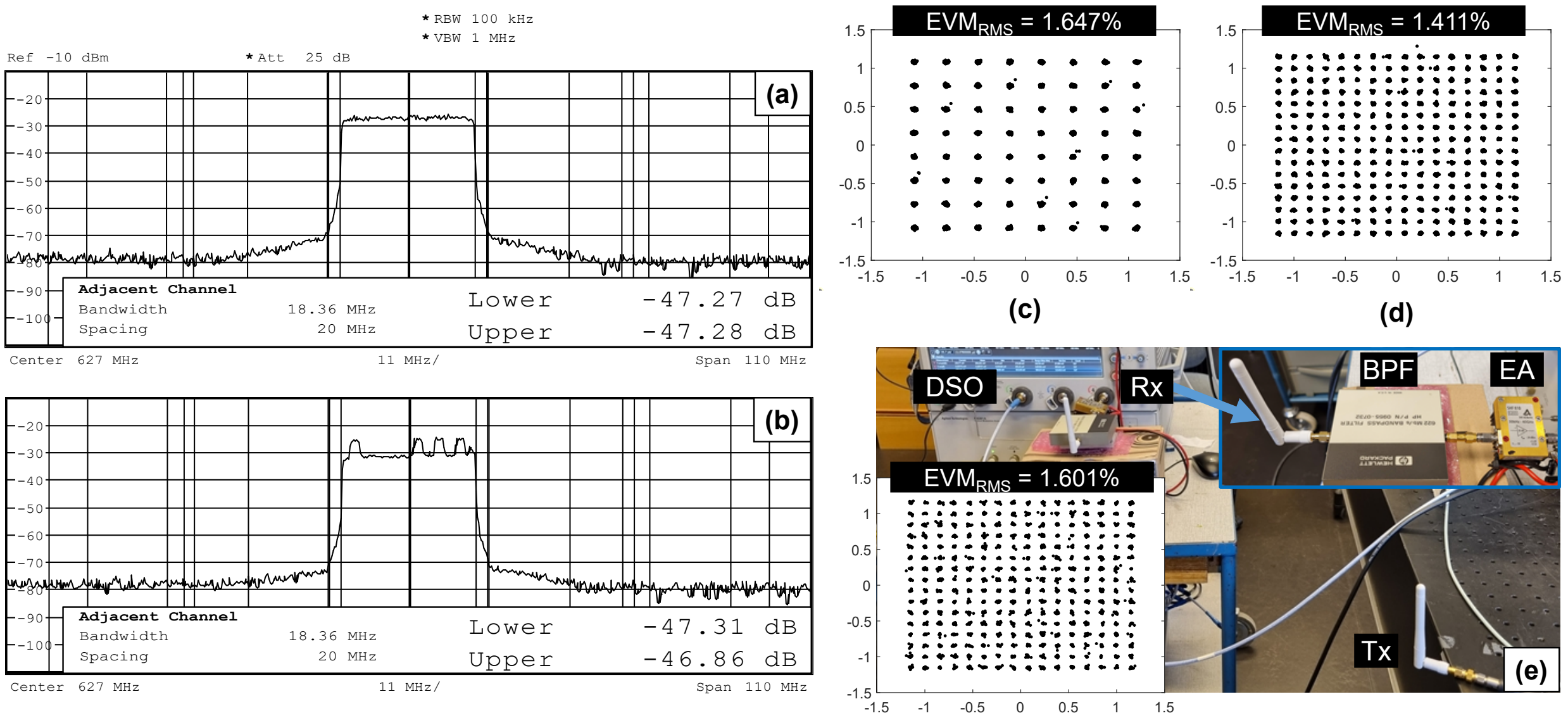

Fig. 2. Electrical spectrum and ACLR measurements of (a) TM NR-FR1-TM1.1 and (b) TM NR-FR1-TM1.2. Received symbols constellation and EVM measurement of (c) TM NR-FR1-TM3.1 and (d) TM NR-FR1-TM3.1a. (e) Photo of wireless transmission setup and received symbols constellation and EVM measurement after wireless transmission of TM NR-FR1-TM3.1a.

Finally, to test the E2E performance, after O/E, we use a multiband antenna to transmit the RF signals wirelessly. At the receiver side, the received signal is first filtered with a narrow BPF centered at 622 MHz and then amplified with a 30 dB gain EA with a 6 dB NF to compensate for the wireless propagation losses. We adjusted our laboratory setup to achieve nearly flat fading conditions on the received signal to realize an additive white Gaussian noise (AWGN) channel. Fig. 2(e) insets show the receiver RF front-end and antenna, and the test signal NR-FR1-TM3.1a received symbols constellation after a wireless transmission of 0.57 m. It can be seen that the EVM degradation is negligible in comparison with the conducted results in Fig. 2(d).

## 4. Conclusion

We have demonstrated, for the first time to our knowledge, LWIR FSO transmission at 9.15 µm of 5G/NR signals fully compliant with the 3GPP RF mandatory EVM and ACLR transmitter requirements. A room-temperature directly-modulated QCL and a commercial MCT detector are used for the LWIR FSO setup. It is noted that, if proper analog and digital RF front-ends with better characteristics, e.g., better NFs, even larger ACLR and EVM margins are expected. Such a conformance test is considered an significant step towards practical analog FSO or hybrid fiber/FSO fronthaul solutions supporting the next-generation D-MIMO network in the upcoming 6G era.

## 5. Acknowledgement

This work is financially supported by the Swedish Foundation for Strategic Research (project No. SM21-0047). Also, this work was supported in part by the EU H2020 cFLOW Project (828893), in part by the Swedish Research Council (VR) projects 2019-05197 and 2016-04510, in part by the by COST Action CA19111 NEWFOCUS, and in part by the ERDF-funded CARAT project (No. 1.1.1.2/VIAA/4/20/660).